\documentclass[11pt,a4paper]{article}
\usepackage{amsmath}
\usepackage{amssymb}
\usepackage{graphicx}
\usepackage{hyperref}
\usepackage[super,sort&compress,comma,square]{natbib}
\usepackage[noblocks]{authblk}
\usepackage[left=2.00cm, right=2.00cm, top=2.00cm, bottom=2.00cm]{geometry}
\usepackage[version=3]{mhchem} 
\usepackage[T1]{fontenc}

\begin{document}

\title{Fast and Highly Sensitive Ionic Polymer Gated \ce{WS2}-Graphene Photodetectors}

\author[1,2,+]{Jake D. Mehew}
\author[1,+]{Selim Unal}
\author[1]{Elias Torres Alonso}
\author[1]{Gareth F. Jones}
\author[1,3]{Saad Fadhil Ramadhan}
\author[1]{Monica F. Craciun}
\author[1,*]{Saverio Russo}
\affil[1]{Centre for Graphene Science, College of Engineering, Mathematics and Physical Sciences, University of Exeter, Exeter, EX4 4QL, UK}
\affil[2]{EPSRC Centre for Doctoral Training in Metamaterials, College of Engineering, Mathematics and Physical Sciences, University of Exeter, Exeter, EX4 4QL, UK}
\affil[3]{Department of Physics, College of Science, University of Duhok, Duhok, 42001 Kurdistan Region, Iraq}
\affil[*]{Email: s.russo@exeter.ac.uk}
\affil[+]{Both authors contributed equally to this work}

\date{}

\maketitle

\setlength{\parskip}{2ex}
\linespread{1}
\selectfont

\begin{abstract}
The combination of graphene with semiconductor materials in heterostructure photodetectors, has enabled amplified detection of femtowatt light signals using micron-scale electronic devices. Presently, the speed of such detectors is limited by long-lived charge traps and impractical strategies, e.g. the use of large gate voltage pulses, have been employed to achieve bandwidths suitable for applications, such as video-frame-rate imaging. Here, we report atomically thin graphene-\ce{WS2} heterostructure photodetectors encapsulated in an ionic polymer, which are uniquely able to operate at bandwidths up to 1.5 kHz, whilst maintaining internal gain as large as $10^6$. Highly mobile ions and a nanometre scale Debye length of the ionic polymer are used to screen charge traps and tune the Fermi level of graphene over an unprecedented range at the interface with \ce{WS2}. We observe a responsivity $R=10^6$ A W$^{-1}$ and detectivity $D^*=3.8\times10^{11}$ Jones, approaching that of single photon counters. The combination of both high responsivity and fast response times makes these photodetectors suitable for video-frame-rate imaging applications.

\end{abstract}


\newpage
The use of two-dimensional (2D) materials in optoelectronic devices has the potential to supersede current state-of-the-art technology\cite{Koppens2014} by added functionalities, such as mechanical flexibility and ease of integration onto textile fibres, enabling the development of new wearable electronic applications.\cite{Neves2015} Graphene transistors have been shown to operate as high-speed photodetectors\cite{Mueller2010} with response times comparable to conventional silicon-based devices, but the absence of a band gap and lack of significant gain mechanism limits its use for ultra-sensitive light detection. Hybrid structures of graphene with semiconductor materials such as quantum dots,\cite{Konstantatos2012,Sun2012d,Nikitskiy2016} chlorophyll molecules,\cite{Chen2013} and \ce{MoS2}\cite{Roy2013a,Roy2013,Zhang2014} have been shown to enhance light absorption and provide an internal gain mechanism. However, these implementations typically have a limited operational bandwidth of less than 10 Hz which hampers their use in real world applications.

Slow response times in these systems are produced by the long-lived trapping of charges, often manifested as hysteresis in gate voltage sweeps. This has been observed in organic, carbon nanotubes, graphene, and more recently in transition metal dichalcogenide (TMD) field-effect transistors, and is typically attributed to unavoidable intrinsic and/or extrinsic charge traps, e.g. \ce{SiO2} surface states\cite{Egginger2009,Joshi2010,Wang2010a,Ghatak2011a} and atmospheric contamination. \cite{Lin2006,Joshi2010,Wang2010a,Late2012,Cho2013} To reduce the impact of such traps, various solutions have been explored including gate voltage pulses,\cite{Egginger2009,Mattmann2010,Ma2015} vacuum annealing,\cite{Yan2011,Ovchinnikov2014} and ionic liquid gating.\cite{Ozel2005,Lin2012} Although ionic liquid gating has been utilised in \ce{WS2} phototransistors\cite{Ubrig2014} and \ce{MoTe2}-graphene photodetectors\cite{Kuiri2016a}, the beneficial effect of polymer gating on the performance of photodetectors consisting of atomically thin heterostructures has not yet been explored.

In this work, we report the first study of \ce{WS2}-graphene heterostructure photodetectors with an ionic polymer gate. We demonstrate a gate tunable responsivity up to $10^6$ A W$^{-1}$, which is comparable with other heterostructure devices,\cite{Konstantatos2012,Sun2012d,Roy2013,Chen2013,Zhang2014,Nikitskiy2016} and surpasses that of graphene or TMD photodetectors by at least 4 orders of magnitude. Our devices reach a -3 dB bandwidth of 1.5 kHz, without the need for any gate pulse, leading to sub-millisecond rise and fall times. The observed $10^3$ fold increase of photodetection bandwidth, when compared to other heterostructure photodetectors, is enabled by the enhanced screening properties of the mobile ions in our ionic polymer top gate, which act to compensate the charge traps limiting the speed of previous devices. Our devices have a detectivity $D^*=3.8\times10^{11}$ Jones, which is approaching that of single photon counters, and are able to operate on a broad spectral range (400 - 700 nm). These properties make ionic polymer gated \ce{WS2}-graphene photodetectors highly suitable for video-frame-rate imaging applications unlike previously developed graphene-based heterostructure photodetectors.\cite{Konstantatos2012,Sun2012d,Chen2013,Roy2013a,Roy2013,Zhang2014}


\begin{figure}[ht]
\centering
\includegraphics[width=0.9\linewidth]{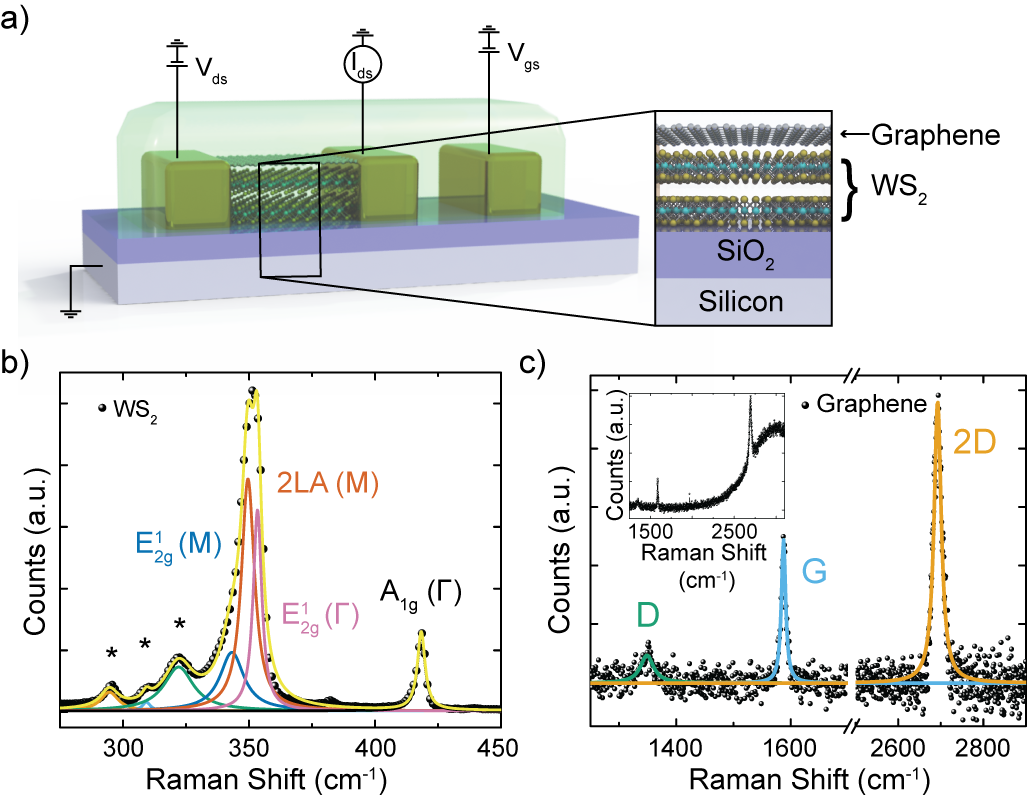}
\caption{
Device schematic and Raman spectrum of the \ce{WS2}-graphene field-effect transistor.
(a) Device schematic with electrical connections included. A voltage ($V_{gs}$) is applied to the transparent polymer (PEO + \ce{LiClO4}) using a gate electrode in close vicinity to the \ce{WS2}-graphene photodetector. Raman spectra of the \ce{WS2}-graphene stack are shown for ranges of wavenumber relevant to (b) \ce{WS2} and (c) graphene. Peaks labelled $*$ are resonant second order processes. Inset in (c) shows the spectrum before baseline subtraction.
}
\label{fig:device}
\end{figure}


Hybrid \ce{WS2}-graphene photodetectors have been fabricated onto p-\ce{Si}/\ce{SiO2} (300 nm) substrates, where the doped \ce{Si} serves as a global back gate. Few-layer \ce{WS2} was mechanically exfoliated from natural crystals and transferred onto the \ce{SiO2} substrate by means of adhesive tape. High quality graphene grown by chemical vapour deposition was then transferred onto the \ce{WS2}, see Supporting Information.\cite{Bointon2015} Electrical contacts were defined by standard electron beam lithography, electron beam deposition of \ce{Au} (20 nm) and lift-off in acetone. Subsequently, conductive graphene channels of widths ranging from 3 to 10 $\mu$m and lengths 1 to 12 $\mu$m were defined by means of \ce{O2} plasma etching. The \ce{WS2}-graphene devices were covered by a transparent ionic polymer, lithium perchlorate/\textit{poly}(ethylene oxide) (\ce{LiClO4}/\ce{PEO}, 8:1 in methanol), which serves as a top gate, see Figure \ref{fig:device}a.

Raman spectroscopy is used for the characterization of \ce{WS2} and graphene heterostructures and reveals peaks in two well-separated regions, 200 cm$^{-1} \le \omega \le$ 450 cm$^{-1}$ and 1200 cm$^{-1} \le \omega \le$ 3000 cm$^{-1}$ respectively. Lorentzian fits of the spectra reveal the presence of several peaks, which originate from the E$_{2g}$, 2LA (second-order longitudinal acoustic) and A$_{1g}$ modes of \ce{WS2} (see Figure \ref{fig:device}b).\cite{Sekine1980} The E$_{2g}$ phonon mode is an in-plane displacement of both sulphur and tungsten atoms, whereas, the A$_{1g}$ mode is an out-of-plane displacement of the sulphur atoms. The position of both modes shifts with increasing numbers of layers, and their wavenumber difference changes with layer number.\cite{Gutierrez2013,Berkdemir2013,Withers2014,Pawbake2016} For the spectra in Figure \ref{fig:device}c, a peak separation of 68.7 cm$^{-1}$ is indicative of a trilayer \ce{WS2} flake. The 2LA peak is a disorder activated overtone of the LA mode, which is the in-plane collective motions of atoms in the lattice.\cite{Berkdemir2013} Resonant enhancement of this mode is observed because the photon energy used in acquiring the Raman spectra lies close to the B exciton energy of \ce{WS2}.\cite{Stacy1985} This is consistent with the broad photoluminescence peak located at $\sim$3100 cm$^{-1}$, attributed to the direct electronic transition of \ce{WS2}. After subtracting this photoluminescence peak from the Raman spectrum, we identify the D, G, and 2D peaks of graphene (Figure \ref{fig:device}c).\cite{Ferrari2013} As we have reported elsewhere these films are monolayer graphene.\cite{Bointon2015} At the same time, the observed low D/G peak intensity ratio ($\sim0.2$) indicates a low defect density.\cite{Bointon2015} Finally, the fact that the measured Raman spectrum on the \ce{WS2}-graphene heterointerface simply is the sum of the individual spectrum for isolated \ce{WS2} and graphene confirms the formation of a van der Waals interface.

\begin{figure}[ht]
	\centering
	\includegraphics[width=0.9\linewidth]{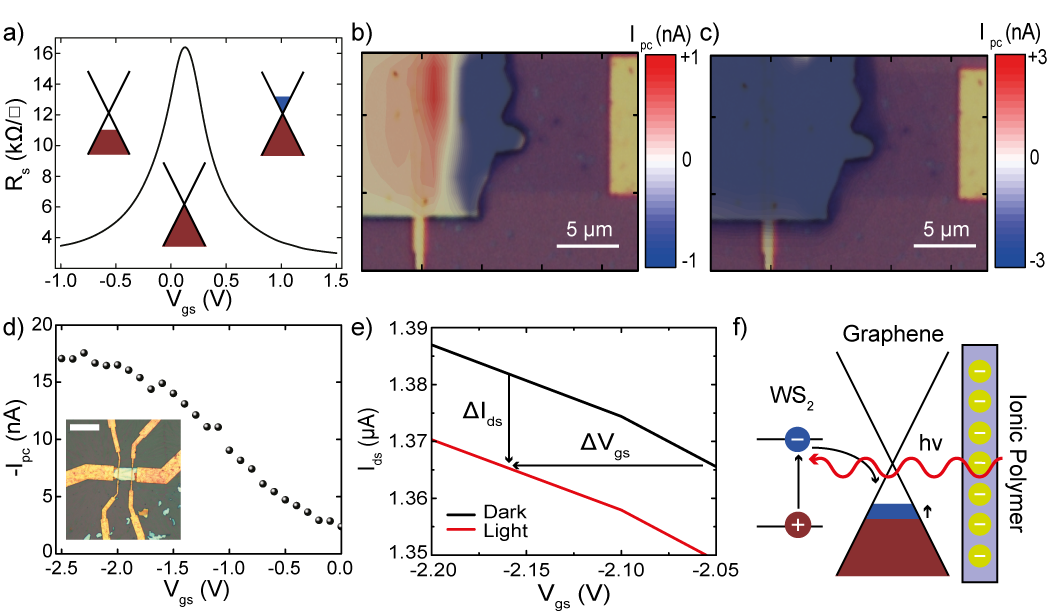}
	\caption{Characterisation of optoelectronic response and charge transfer mechanism. 
	(a) Channel resistance ($R_s$) as a function of gate voltage ($V_{gs}$). Insets show schematics of Fermi level position.
	Scanning photocurrent maps of a large area device in short circuit configuration, ($V_{ds} = 0$ mV), (b) and under a source drain bias, ($V_{ds} = 5$ mV), (c). 
	(d) Photocurrent ($I_{pc}$) versus top gate voltage ($V_{gs}$). Inset shows optical micrograph of device. Scale bar is 16 $\mu$m.
	(e) Drain current ($I_{ds}$) versus $V_{gs}$ in dark and under illumination.
	(f) Schematic of charge transfer at \ce{WS2}/graphene interface.
	}
	\label{fig:origin}
\end{figure}

Figure \ref{fig:origin}a shows the typical ambipolar electrical transport of graphene. Upon applying a bias to the ionic polymer a stable electric double layer is formed at the interface with graphene without the occurrence of chemical reactions within the electrochemical stability window, -2 V $< V_{gs} <$ 2 V. An extremely large gate capacitance easily attained in ionic gated transistors ($\ge2\,\times\,10^{-6}$ F cm$^{-2}$) allows us to probe the properties of graphene at record high charge carrier densities $\ge10^{14}$ cm$^{-2}$.\cite{Ye2011,Lu2004} Most importantly, the ions in the polymer are highly mobile and provide a significant additional screening mechanism of charge impurities.\cite{Sharma2015}

To determine the photo-responsive region of the fabricated \ce{WS2}-graphene hybrid structures we use Scanning Photocurrent Microscopy (SPCM) which employs a focussed laser beam, see Experimental Section.\cite{DeSanctis2016} Figure \ref{fig:origin}b shows that in the short circuit configuration ($V_{ds} = 0$ V) photocurrent generation is localised to the lateral interfaces of the device, such as the edges of Au contacts and the \ce{WS2} flake, and changes in polarity across the  photo-responsive region. Upon applying a finite source-drain bias, a uniform photocurrent is generated over the entire vertical \ce{WS2}-graphene interface, see Figure \ref{fig:origin}c.

To gain insight in the microscopic origin of the measured photocurrent and understand the role played by the ionic polymer gate on device performance, we characterize the photoresponse of these structures in a vacuum chamber at finite source-drain bias and under illumination with collimated light (see Experimental Section). Figure \ref{fig:origin}d shows that upon increasing top gate voltage ($V_{gs}$) $I_{pc}$ increases until $V_{gs} = -2$ V, at which point $I_{pc}$ reaches a peak value of $-18$ nA. For $V_{gs} \le -2$ V no further increase in $I_{pc}$ is observed. To explain the increased photocurrent under a gate bias we examine the transfer curves (V$_{ds}$ = 10 mV) taken in both dark and light (600 nm, 200 $\mu$W cm$^{-2}$) conditions as seen in Figure \ref{fig:origin}e. Under illumination a reduction in the current ($\Delta I_{ds}$) is observed and this increases for more negative gate biases. This is expected when the photocurrent generation mechanism is the photogating effect\cite{Koppens2014} where absorption of photons in \ce{WS2} creates electron-hole pairs, which can be split at the interface between graphene and \ce{WS2}, with one charge carrier transferred to graphene and the other remaining in \ce{WS2}, as shown schematically in Figure \ref{fig:origin}f. The in-built fields at the interface enable this separation and arise from the work function difference between graphene and \ce{WS2}. For $V_{gs} < V_{dirac}$ illumination of the heterostructure results in an increase in resistance due to the recombination between electrons, generated in \ce{WS2} and subsequently transferred to graphene, and electrostatically induced holes present in graphene. This manifests as a shift in the charge neutrality point ($\Delta V_{gs}$) to negative values, indicating n-type doping. Photogenerated holes remain trapped in \ce{WS2} and could be considered as a light induced gating potential.

\begin{figure}[ht]
	\centering
	\includegraphics[width=0.9\linewidth]{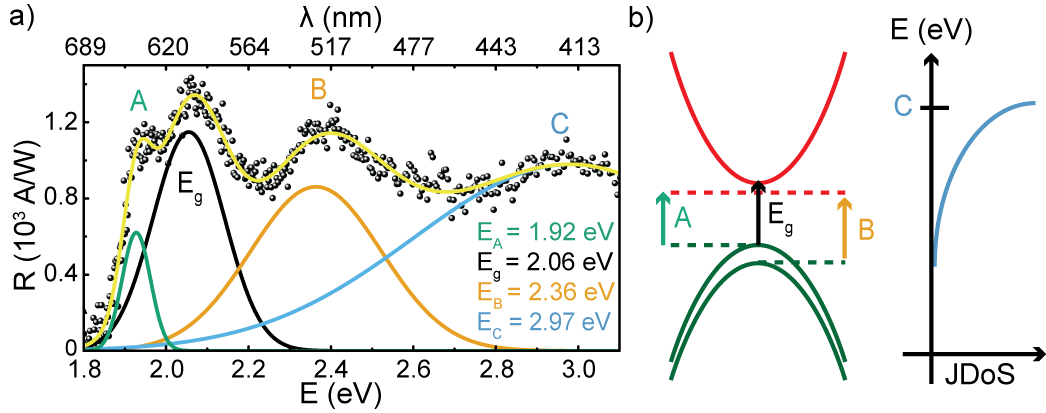}
	\caption{Characterisation of the spectral response of \ce{WS2}-graphene heterostructures.
	(a) Measured responsivity (black dots) versus incident photon energy for $\text{V}_{ds}=10 \text{ mV}$ and $\text{V}_{gs}=0 \text{ V}$. A, B and C exciton peaks, as well as the direct gap transition ($E_g$), are fit with Gaussian functions with the cumulative fit described by the yellow continuous line.
	(b) Schematic of electronic transitions responsible for each peak fitted in the spectral responsivity of \ce{WS2}/graphene interface.
	}
	\label{fig:spec}
\end{figure}

These devices display an energy dependent responsivity ($R$) when illuminated by monochromatic light, see Figure \ref{fig:spec}a. More specifically, a photoresponse is only observed for incident photons of energy greater than 1.8 eV, with the spectral profile of responsivity consisting of four Gaussian peaks centred at 1.92 eV, 2.06 eV and 2.36 eV, with a broader peak at 2.97 eV also present. All of these peaks relate to different electronic transitions in \ce{WS2}, as illustrated in Figure \ref{fig:spec}b. The peak at 2.06 eV is the single particle band gap, E$_g$, and at 1.92 eV we also observe the peak arising from the A exciton.\cite{Wilson1969} This exciton corresponds to the electronic transition from the upper branch of the split valence band to the conduction band, and subsequent formation of a bound state between an electron and hole. 

In most semiconductors excitons can be described using a Wannier-Mott 2D hydrogen model.\cite{Chernikov2014} Although the applicability of this model to 2D systems can be questioned because of the increased exciton confinement and reduced electric field screening,\cite{Chernikov2014} in this case we find that it serves as a reasonable approximation. From the model we can extract the binding energy, $\beta$, using $E_A = E_g - \beta$ which gives $\beta = 140$ meV which is a value between that of bulk ($\sim50$ meV)\cite{Wilson1969} and monolayer ($\sim300 - 800$ meV)\cite{Chernikov2014,Hanbicki2015} \ce{WS2}. Such a high binding energy inhibits the contribution of excitons to the measured photocurrent unless they can dissociate into an unbound electron-hole pair and be transferred to the graphene charge transport layer.\cite{Klots2014} This dissociation can occur as long as the binding energy can be overcome which typically requires large electric fields. The in-built field at the interface, arising from the work function mismatch ($\Delta\phi$) between graphene and \ce{WS2} could encourage this dissociation, although our estimate of $\Delta\phi \sim 100$ meV indicates that this alone would not be sufficient. Applying a non-zero value of $V_{gs}$ creates large electric fields at the surface of graphene which can contribute to the exciton dissociation in \ce{WS2} as the fields are not completely screened by graphene.\cite{Li2015} This has been verified by taking spectral scans at different top gate biases (see Figure S1). 

Finally, the peak at 2.36 eV is due to the exciton formed from the electronic transition originating in the lower branch of the valence band. The difference in energy between this B exciton and the A exciton allows us to extract a spin-orbit splitting energy of $440$ meV, which is in good agreement with both theoretical \cite{Ramasubramaniam2012} and other experimental \cite{Zhou2013} works. The broad peak at 2.97 eV, Figure \ref{fig:spec}a, can be attributed to transitions between regions of high density of states in the valence and conduction bands which give these materials their strong light-matter interaction.\cite{Britnell2013} The joint density of states (JDoS) exhibits this in a clearer fashion and has a prominent peak around this energy, see Figure \ref{fig:spec}b.

\begin{figure}[ht]
	\centering
	\includegraphics[width=0.9\linewidth]{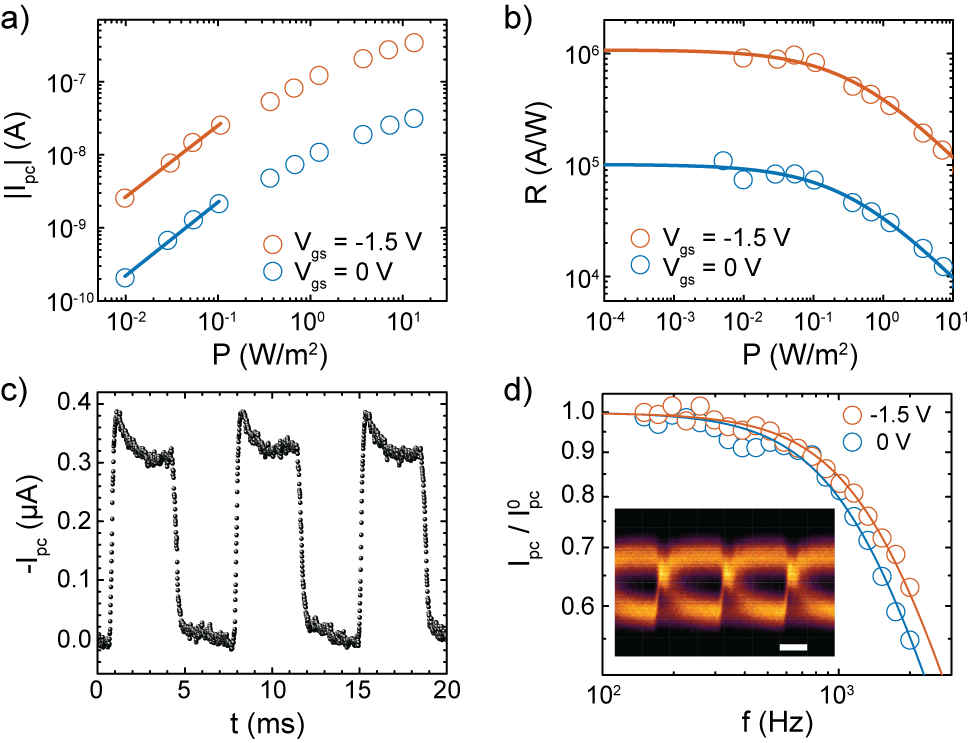}
	\caption{Photodetector performance of \ce{WS2}-graphene heterostructures.
	(a) Photocurrent ($|I_{pc}|$) and (b) responsivity (R) as a function of incident optical power (P) at $V_{ds} = 100$ mV. 
	(c) Temporal response of the device at $V_{ds} = 100$ mV and $V_{gs} = -1.5$ V.
	(d) Normalised photoresponse as a function of light modulation frequency. Inset shows eye diagram acquired at 2.9 kbit s$^{-1}$. Scale bar is 150 $\mu$s.
	}
	\label{fig:perform}
\end{figure}

Hence, in our devices the whole \ce{WS2}-graphene interface is photoactive and its photosensitivity extends across the spectral range 400 - 700 nm. To fully characterise the device performance we have illuminated the device with monochromatic light ($\lambda = 625$ nm) of varying intensity and the photocurrent was recorded. Figure \ref{fig:perform}a shows the photocurrent as a function of incident optical power at zero and finite negative bias applied to the polymer gate. A maximum photocurrent of 339 nA is recorded for an incident power density of approximately 15 W m$^{-2}$, which reduces to 2.55 nA at the lowest detectable illumination levels (V$_{gs}$ = -1.5 V, V$_{ds}$ = 100 mV. For both gate voltages the photocurrent decreases with reducing optical power, transitioning from a sub-linear power dependence to a linear one below $0.1$ W/m$^2$. In the linear regime, indicated by the straight line fits, photogenerated charge carriers are split, with one charge type being transferred to the graphene channel whilst the other remains trapped in the \ce{WS2}. Upon increasing the illumination intensity, the large number of photogenerated charge carriers reduces the electric field at the heterointerface, resulting in a sub-linear power dependence.\cite{Konstantatos2012,Roy2013} Application of a bias to the polymer gate allows for more efficient exciton splitting within \ce{WS2} leading to an increase in $I_{pc}$, as seen previously in Figure \ref{fig:origin}d.

In Figure \ref{fig:perform}b we plot the responsivity as a function of incident optical power for both $V_{gs} = 0$ V and $V_{gs} = -1.5$ V. The responsivity has been calculated using $R = I_{pc}/P$, where $I_{pc}$ is the photocurrent and $P$ the incident optical power, and follows a non-linear power dependence. This can be well fit using a function of the form $R = a/(b+P^n)$, where $a$, $b$ and $n$ are fitting parameters. The power exponent $n\sim2/3$ is indicative of non-radiative Auger recombination,\cite{Hall1959} previously observed in other indirect semiconductors such as Ge and Si.\cite{Wagner1996} In our devices the responsivities reach a maximal value of $1\times10^6$ A W$^{-1}$ at $V_{gs} = -1.5$ V for $V_{ds} = 100$ mV, an order of magnitude higher than that without a bias applied to the top gate, corresponding to an external quantum efficiency of 2.0$\times10^6$. The high responsivities observed in these devices can be explained in terms of a gain mechanism arising from the aforementioned photogating effect; to maintain charge conservation the removal of one electron at a contact requires the injection of one at the opposite contact. This electron circulation exists as long as the holes remain trapped in the \ce{WS2} resulting in a net gain.\cite{Koppens2014} The gain ($G$) in our devices can be theoretically calculated considering the change in carrier density ($\Delta n$) from a known photon flux ($\phi$), see Supporting Information. This gives a value of $G_{th} = 3.4\times10^6$, which is in excellent agreement with our experimental measurement of responsivity, Figure \ref{fig:perform}b.

The temporal response of a polymer gated \ce{WS2}-graphene device is shown in Figure \ref{fig:perform}c at $V_{ds} = 100$ mV and $V_{gs} = -1.5$ V whilst the incident light is modulated at 140 Hz. The rise and fall times are defined as the time period taken for $\Delta I_{pc}$ to change from 10\,\% (90\,\%) to 90\,\% (10\,\%) of its maximum value respectively. Analysing multiple iterations of this square wave signal, we find to high precision that the transient response of the \ce{WS2}-graphene photodetectors takes place over sub-millisecond timescales with $\tau_{\text{rise}} = 130$ $\mu$s and $\tau_{\text{fall}} = 440$  $\mu$s. Prior to encapsulation in the ionic polymer our devices typically had rise and fall times >1 s (see Figure S2), often with the decay of the photocurrent signal persisting well beyond the time frame of the experiment. After deposition of the ionic polymer the response times of our devices improved by at least four orders of magnitude, resulting in sub-millisecond rise and fall times, as seen in Figure \ref{fig:perform}c. 

These response times are $10^4$ faster than previously reported heterostructure photodetectors which utilise TMDs\cite{Roy2013,Zhang2014} or QDs\cite{Konstantatos2012,Sun2012d} as a light absorbing layer, typically operating over time scales of seconds or greater, owing to long lived charge trapping present in these devices. Typically, a large gate pulse is applied to reduce the potential barrier between graphene and the semiconductor, thereby accelerating the recombination rate of photogenerated electrons and holes, allowing a swift transition back to dark conditions. Indeed, hysteresis in current-gate sweeps of carbon nanotubes, attributed to atmospheric contamination and oxide charge traps, can be resolved through gate pulsing strategies.\cite{Mattmann2010} However, for graphene-QDs these gate-pulses have been found to be device specific.\cite{Konstantatos2012} Our devices exhibit rise and fall times that are up to five orders of magnitude faster than these previous works, without the need to apply large electrical pulses. This surprising finding is the result of the ability of mobile ions in the polymer electrolyte to efficiently screen charge traps responsible for the localization of charge carriers in monolayer TMDs.\cite{Ghatak2011a} To date the screening properties of polymer electrolytes have been widely demonstrated in electrical transport measurements.\cite{Ye2011} Here, we harvest this aspect of fundamental physics to reduce the role of long-lived trap states in atomically thin photodetectors, demonstrating an unprecedented fast time response without the need for any gate voltage pulsing strategies

In Figure \ref{fig:perform}d, we verify these response times by ascertaining the -3 dB bandwidth of polymer gated \ce{WS2}-graphene photodetectors by measuring the decline in photocurrent magnitude as an incident light signal is modulated with increasing frequency, using an optical chopper wheel. A similar trend is shown for the situation with and without a bias applied to the polymer gate, where photocurrent signals are normalised to the maximum, which occurs at low modulation frequencies. The normalised signal reduces when increasing frequency, as one would expect when the period of modulation begins to impinge upon the rise and fall times of the device. The -3 dB bandwidth, a common figure of merit for photodetectors, is the point at which the signal has dropped to 70\% of its initial value, which for our devices are 1.3 kHz ($V_{gs} = 0$ V) and 1.5 kHz ($V_{gs} = -1.5$ V). From this we can extract a rise time using $\tau_{rise} \approx 0.35/f_{-3dB}$ of 220 $\mu$s, in good agreement with the data extracted from Figure \ref{fig:perform}c.

This -3 dB bandwidth of 1.5 kHz, coupled with extremely sensitive photodetection across a broad spectral range, means that \ce{WS2}-graphene heterostructures are highly suitable for video-frame-rate imaging applications, thanks to the unique screening properties of the ionic polymer top gate. To demonstrate the feasibility of this claim we constructed a home-built optical data link, with a pseudo-random bit sequence generator used to modulate the 625 nm light of a light-emitting diode. This light was focussed onto the \ce{WS2}-graphene heterostructure maintained at $V_{ds} = 100\text{ mV}$ and $V_{gs} = -1.5\text{ V}$ and the output data stream amplified and delivered into an oscilloscope to obtain an eye diagram. The inset in Figure \ref{fig:perform}d shows such an eye diagram, with the open eye at 2.9 kbit s$^{-1}$ demonstrating that these heterostructures can truly be used in video-frame-rate imaging applications. Our polymer electrolyte encapsulated photodetectors exhibit a gain-bandwidth product of 7.2 GHz which is comparable to established technology based on III-V phototransistors, see Supporting Information.\cite{Leu1991}

Finally, in order to compare the performance of \ce{WS2}-graphene heterostructures to that of other photodetectors, we use the specific detectivity ($D^*$). This can be calculated using the responsivity ($R$) and the noise density ($S_n$, see Figure S3) using $D^* = R \sqrt{A}/S_n$ where A is the device area. Taking the responsivity at $V_{gs} = -1.5$ V and noise value extracted at 150 Hz, we calculate a $D^*=3.8\times10^{11} $ Jones which is comparable to other graphene hybrid photodetectors.\cite{Konstantatos2012,Sun2012d,Roy2013,Chen2013,Zhang2014,Nikitskiy2016,Tan2016a}

To summarise, we have characterised the optoelectronic properties of ionic polymer gated \ce{WS2}-graphene heterostructure photodetectors across a broad spectral range. The photogating effect has been found to be the dominant photocurrent generation mechanism, with a high gain process resulting in responsivities of $1\times10^6$ A W$^{-1}$. Furthermore, we demonstrate sub-millisecond response times of our devices through both rise and fall time estimates as well as by measuring a -3 dB bandwidth of 1.5 kHz. The high gain and fast response found in our devices arises from the ability to compensate charge traps with the ionic polymer, which is a limiting factor in similar photodetectors. Our study demonstrates that both high gain and sub-millisecond response times can be achieved in two-dimensional heterostructure photodetectors. A calculated detectivity of $3.8\times10^{11}$ Jones brings the realisation of high frame-rate video-imaging applications with 2D materials ever closer.

\subsection*{Experimental Section}
\textbf{Optoelectronic measurements.}
Raman spectra were acquired using a 532 nm laser source with a spot size of $\sim1$ $\mu$m and an incident laser beam power <40 $\mu$W to avoid overheating and damage to \ce{WS2}-graphene. Photocurrent maps were recorded at room temperature in ambient conditions in a custom built set-up on an upright BX51 Olympus microscope described and characterised extensively in reference.\cite{DeSanctis2016} The external quantum efficiency, spectral responsivity, and transient response measurements were performed in a custom built vacuum chamber (10$^{-3}$ mbar) using a Xenon Lamp, monochromator and collimating optics (Oriel TLS-300X), to provide a spectrally tunable incident light source. Neutral density filters and a motorized chopper wheel were used to attenuate and modulate the incident signal respectively. Power calibrations were performed with a ThorLabs PM320E power meter equipped with a S130VC sensor.

\subsection*{Supporting Information}
Supporting Information is available from \url{http://doi.org/10.1002/adma.201700222} or from the author.

\subsection*{Acknowledgements}

J.D.M. and S.U. contributed equally to this work. The authors thank Adolfo De Sanctis for assistance in obtaining the photocurrent maps and useful discussions. J.D.M. acknowledges financial support from the Engineering and Physical Sciences Research Council (EPSRC) of the United Kingdom, via the EPSRC Centre for Doctoral Training in Metamaterials (Grant No. EP/L015331/1 ). S.F.R acknowledges financial support from the Higher Committee for Education Development in Iraq (HCED). S.R. and M.F.C. acknowledge financial support from EPSRC (Grant No. EP/J000396/1, EP/K017160/1, EP/K010050/1, EP/ G036101/1, EP/M001024/1, and EP/M002438/1) and from Royal Society International Exchanges Scheme 2016/R1.


\begin{thebibliography}{10}

\bibitem{Koppens2014}
F.~H. Koppens, T.~Mueller, P.~Avouris, a.~C. Ferrari, M.~S. Vitiello,
  M.~Polini, \emph{Nat Nanotechnol} \textbf{2014}, \emph{9}, 780.

\bibitem{Neves2015}
A.~I.~S. Neves, T.~H. Bointon, L.~V. Melo, S.~Russo, I.~de~Schrijver, M.~F.
  Craciun, H.~Alves, \emph{Scientific Reports} \textbf{2015}, \emph{5}, 9866.

\bibitem{Mueller2010}
T.~Mueller, F.~Xia, P.~Avouris, \emph{Nature Photonics} \textbf{2010},
  \emph{4}, 297.

\bibitem{Konstantatos2012}
G.~Konstantatos, M.~Badioli, L.~Gaudreau, J.~Osmond, M.~Bernechea, F.~P.~G.
  de~Arquer, F.~Gatti, F.~H.~L. Koppens, \emph{Nature Nanotechnology}
  \textbf{2012}, \emph{7}, 363.

\bibitem{Sun2012d}
Z.~Sun, Z.~Liu, J.~Li, G.-a. Tai, S.-P. Lau, F.~Yan, \emph{Advanced Materials}
  \textbf{2012}, \emph{24}, 5878.

\bibitem{Nikitskiy2016}
I.~Nikitskiy, S.~Goossens, D.~Kufer, T.~Lasanta, G.~Navickaite, F.~H.~L.
  Koppens, G.~Konstantatos, \emph{Nature Communications} \textbf{2016},
  \emph{7}, 11954.

\bibitem{Chen2013}
S.~Y. Chen, Y.~Y. Lu, F.~Y. Shih, P.~H. Ho, Y.~F. Chen, C.~W. Chen, Y.~T. Chen,
  W.~H. Wang, \emph{Carbon} \textbf{2013}, \emph{63}, 23.

\bibitem{Roy2013a}
K.~Roy, M.~Padmanabhan, S.~Goswami, T.~P. Sai, S.~Kaushal, A.~Ghosh,
  \emph{Solid State Communications} \textbf{2013}, \emph{175-176}, 35.

\bibitem{Roy2013}
K.~Roy, M.~Padmanabhan, S.~Goswami, T.~P. Sai, G.~Ramalingam, S.~Raghavan,
  A.~Ghosh, \emph{Nature Nanotechnology} \textbf{2013}, \emph{8}, 826.

\bibitem{Zhang2014}
W.~Zhang, C.-P. Chuu, J.-K. Huang, C.-H. Chen, M.-L. Tsai, Y.-H. Chang, C.-T.
  Liang, Y.-Z. Chen, Y.-L. Chueh, J.-H. He, M.-Y. Chou, L.-J. Li,
  \emph{Scientific reports} \textbf{2014}, \emph{4}, 3826.

\bibitem{Egginger2009}
M.~Egginger, S.~Bauer, R.~Schw{\"{o}}diauer, H.~Neugebauer, N.~S. Sariciftci,
  \emph{Monatshefte fur Chemie} \textbf{2009}, \emph{140}, 735.

\bibitem{Joshi2010}
P.~Joshi, H.~E. Romero, a.~T. Neal, V.~K. Toutam, S.~a. Tadigadapa,
  \emph{Journal of physics. Condensed matter : an Institute of Physics journal}
  \textbf{2010}, \emph{22}, 334214.

\bibitem{Wang2010a}
H.~Wang, Y.~Wu, C.~Cong, J.~Shang, T.~Yu, \emph{ACS Nano} \textbf{2010},
  \emph{4}, 7221.

\bibitem{Ghatak2011a}
S.~Ghatak, A.~N. Pal, A.~Ghosh, \emph{ACS Nano} \textbf{2011}, \emph{5}, 7707.

\bibitem{Lin2006}
H.~Lin, S.~Tiwari, \emph{Applied Physics Letters} \textbf{2006}, \emph{89},
  073507.

\bibitem{Late2012}
D.~J. Late, B.~Liu, H.~S. S.~R. Matte, V.~P. Dravid, C.~N.~R. Rao, \emph{ACS
  Nano} \textbf{2012}, \emph{6}, 5635.

\bibitem{Cho2013}
K.~Cho, W.~Park, J.~Park, H.~Jeong, J.~Jang, T.~Y. Kim, W.~K. Hong, S.~Hong,
  T.~Lee, \emph{ACS Nano} \textbf{2013}, \emph{7}, 7751.

\bibitem{Mattmann2010}
M.~Mattmann, C.~Roman, T.~Helbling, D.~Bechstein, L.~Durrer, R.~Pohle,
  M.~Fleischer, C.~Hierold, \emph{Nanotechnology} \textbf{2010}, \emph{21},
  185501.

\bibitem{Ma2015}
C.~Ma, Y.~Gong, R.~Lu, E.~Brown, B.~Ma, J.~Li, J.~Wu, \emph{Nanoscale}
  \textbf{2015}, \emph{7}, 18489.

\bibitem{Yan2011}
J.~Yan, M.~S. Fuhrer, \emph{Physical Review Letters} \textbf{2011}, \emph{107},
  206601.

\bibitem{Ovchinnikov2014}
D.~Ovchinnikov, A.~Allain, Y.~S. Huang, D.~Dumcenco, A.~Kis, \emph{ACS Nano}
  \textbf{2014}, \emph{8}, 8174.

\bibitem{Ozel2005}
T.~Ozel, A.~Gaur, J.~A. Rogers, M.~Shim, \emph{Nano Letters} \textbf{2005},
  \emph{5}, 905.

\bibitem{Lin2012}
M.-W. Lin, L.~Liu, Q.~Lan, X.~Tan, K.~S. Dhindsa, P.~Zeng, V.~M. Naik, M.~M.-C.
  Cheng, Z.~Zhou, \emph{Journal of Physics D: Applied Physics} \textbf{2012},
  \emph{45}, 345102.

\bibitem{Ubrig2014}
N.~Ubrig, S.~Jo, H.~Berger, A.~F. Morpurgo, A.~B. Kuzmenko, \emph{Applied
  Physics Letters} \textbf{2014}, \emph{104}, 171112.

\bibitem{Kuiri2016a}
M.~Kuiri, B.~Chakraborty, A.~Paul, S.~Das, A.~K. Sood, A.~Das, \emph{Applied
  Physics Letters} \textbf{2016}, \emph{108}, 063506.

\bibitem{Bointon2015}
T.~H. Bointon, M.~D. Barnes, S.~Russo, M.~F. Craciun, \emph{Advanced materials
  (Deerfield Beach, Fla.)} \textbf{2015}, \emph{27}, 4200.

\bibitem{Sekine1980}
T.~Sekine, T.~Nakashizu, K.~Toyoda, K.~Uchinokura, E.~Matsuura, \emph{Solid
  State Communications} \textbf{1980}, \emph{35}, 371.

\bibitem{Gutierrez2013}
H.~R. Guti{\'{e}}rrez, N.~Perea-L{\'{o}}pez, A.~L. El{\'{i}}as, A.~Berkdemir,
  B.~Wang, R.~Lv, F.~L{\'{o}}pez-Ur{\'{i}}as, V.~H. Crespi, H.~Terrones,
  M.~Terrones, \emph{Nano Letters} \textbf{2013}, \emph{13}, 3447.

\bibitem{Berkdemir2013}
A.~Berkdemir, H.~R. Guti{\'{e}}rrez, A.~R. Botello-M{\'{e}}ndez,
  N.~Perea-L{\'{o}}pez, A.~L. El{\'{i}}as, C.-I. Chia, B.~Wang, V.~H. Crespi,
  F.~L{\'{o}}pez-Ur{\'{i}}as, J.-C. Charlier, H.~Terrones, M.~Terrones,
  \emph{Scientific Reports} \textbf{2013}, \emph{3}, 1755.

\bibitem{Withers2014}
F.~Withers, T.~H. Bointon, D.~C. Hudson, M.~F. Craciun, S.~Russo,
  \emph{Scientific Reports} \textbf{2014}, \emph{4}, 4967.

\bibitem{Pawbake2016}
A.~S. Pawbake, M.~S. Pawar, S.~R. Jadkar, D.~J. Late, \emph{Nanoscale}
  \textbf{2016}, \emph{8}, 3008.

\bibitem{Stacy1985}
A.~M. Stacy, D.~T. Hodul, \emph{Journal of Physics and Chemistry of Solids}
  \textbf{1985}, \emph{46}, 405.

\bibitem{Ferrari2013}
A.~C. A.~C. Ferrari, D.~M.~D. Basko, \emph{Nature Nanotechnology}
  \textbf{2013}, \emph{8}, 235.

\bibitem{Ye2011}
J.~Ye, M.~F. Craciun, M.~Koshino, S.~Russo, S.~Inoue, H.~Yuan, H.~Shimotani,
  A.~F. Morpurgo, Y.~Iwasa, \emph{Proceedings of the National Academy of
  Sciences of the United States of America} \textbf{2011}, \emph{108}, 13002.

\bibitem{Lu2004}
C.~Lu, Q.~Fu, S.~Huang, J.~Liu, \emph{Nano Letters} \textbf{2004}, \emph{4},
  623.

\bibitem{Sharma2015}
P.~Sharma, Z.~L. Mi{\v{s}}kovi{\'{c}}, \emph{The Journal of Chemical Physics}
  \textbf{2015}, \emph{143}, 134118.

\bibitem{DeSanctis2016}
A.~{De Sanctis}, G.~F. Jones, N.~J. Townsend, M.~F. Craciun, S.~Russo,
  \emph{arXiv} \textbf{2016}.

\bibitem{Wilson1969}
J.~Wilson, A.~Yoffe, \emph{Advances in Physics} \textbf{1969}, \emph{18}, 193.

\bibitem{Chernikov2014}
A.~Chernikov, T.~C. Berkelbach, H.~M. Hill, A.~Rigosi, Y.~Li, O.~B. Aslan,
  D.~R. Reichman, M.~S. Hybertsen, T.~F. Heinz, \emph{Physical Review Letters}
  \textbf{2014}, \emph{113}, 1.

\bibitem{Hanbicki2015}
A.~T. Hanbicki, M.~Currie, G.~Kioseoglou, A.~L. Friedman, B.~T. Jonker,
  \emph{Solid State Communications} \textbf{2015}, \emph{203}, 16.

\bibitem{Klots2014}
A.~R. Klots, A.~K.~M. Newaz, B.~Wang, D.~Prasai, H.~Krzyzanowska, J.~Lin,
  D.~Caudel, N.~J. Ghimire, J.~Yan, B.~L. Ivanov, K.~A. Velizhanin, A.~Burger,
  D.~G. Mandrus, N.~H. Tolk, S.~T. Pantelides, K.~I. Bolotin, \emph{Scientific
  Reports} \textbf{2014}, \emph{4}, 6608.

\bibitem{Li2015}
Y.~Li, C.~Y. Xu, J.~K. Qin, W.~Feng, J.~Y. Wang, S.~Zhang, L.~P. Ma, J.~Cao,
  P.~A. Hu, W.~Ren, L.~Zhen, \emph{Advanced Functional Materials}
  \textbf{2016}, \emph{26}, 293.

\bibitem{Ramasubramaniam2012}
A.~Ramasubramaniam, \emph{Physical Review B - Condensed Matter and Materials
  Physics} \textbf{2012}, \emph{86}, 115409.

\bibitem{Zhou2013}
W.~Zhao, Z.~Ghorannevis, L.~Chu, M.~Toh, C.~Kloc, P.-H.~H. Tan, G.~Eda,
  \emph{ACS Nano} \textbf{2013}, \emph{7}, 791.

\bibitem{Britnell2013}
L.~Britnell, R.~M. Ribeiro, A.~Eckmann, R.~Jalil, B.~D. Belle, A.~Mishchenko,
  Y.~Kim, R.~V. Gorbachev, T.~Georgiou, S.~V. Morozov, A.~N. Grigorenko, A.~K.
  Geim, C.~Casiraghi, A.~H.~C. Neto, K.~S. Novoselov, \emph{Science}
  \textbf{2013}, \emph{340}, 1331.

\bibitem{Hall1959}
R.~Hall, \emph{Proceedings of the IEE - Part B: Electronic and Communication
  Engineering} \textbf{1959}, \emph{106}, 923.

\bibitem{Wagner1996}
R.~E. Wagner, A.~Mandelis, \emph{Semiconductor Science and Technology}
  \textbf{1996}, \emph{11}, 300.

\bibitem{Leu1991}
L.~Y. Leu, J.~T. Gardner, S.~R. Forrest, \emph{Journal of Applied Physics}
  \textbf{1991}, \emph{69}, 1052.

\bibitem{Tan2016a}
H.~Tan, Y.~Fan, Y.~Zhou, Q.~Chen, W.~Xu, J.~H. Warner, \emph{ACS Nano}
  \textbf{2016}, \emph{10}, 7866.

\end{thebibliography}
\end{document}